\documentclass[prd,twocolumn,floatfix,preprintnumbers,letterpaper]{revtex4}
\usepackage{graphics}
\usepackage{epsfig}
\usepackage{amsmath}

\def\4he{$^4$He}

\begin{document}

\title{Big Bang Nucleosynthesis Constraints on Brane Cosmologies}
\author{Jonathan D. Bratt}
\affiliation{Department of Physics,
Geneva College, Beaver Falls, PA ~15010}
\author{A.C. Gault}
\affiliation{Department of Physics and Astronomy, The University of Toledo, Toledo, OH~~43606}
\author{Robert J. Scherrer}
\affiliation{Department of Physics and Department of Astronomy, 
The Ohio State University,
Columbus, OH~~43210}
\author{T.P. Walker}
\affiliation{Department of Physics and Department of Astronomy, 
The Ohio State University,
Columbus, OH~~43210}

\begin{abstract}
We examine constraints from Big Bang nucleosynthesis on type II
Randall-Sundrum brane cosmologies with
both a dark radiation component and a quadratic term
that depends on the 5-dimensional Planck mass, $M_5$.  Using
limits on the abundances of deuterium and helium-4, we calculate
the allowed region in the $M_5$--dark radiation plane and derive
the precise BBN bound on $M_5$ alone with no dark
radiation:  $M_5 > 13$ TeV.

\end{abstract}

\maketitle

In brane cosmologies, the observable universe is a brane embedded in
a higher-dimensional bulk.  The standard-model fields are confined to our
3-brane, while gravity alone propagates in the bulk.
Such models were proposed to explain the large gap between the
energy scale of gravity and the energy scales of the other interactions;
by adding a number of extra compact dimensions, the ``true" Planck
mass can be moved down to the TeV scale \cite{Arkani,Anton}.

An interesting variation on this idea was proposed by Randall and Sundrum
\cite{Randall}, who produced a set of models with a non-compactified extra dimension.
In the type II Randall-Sundrum model, the brane has positive tension, and
the bulk contains a negative cosmological constant.  The cosmology
produced in this model has been investigated in some detail
\cite{Csaki,Cline,Binetruy}.  With the correct
fine-tuning, the Friedmann equation in this scenario reduces to:
\begin{equation}
\label{Friedmann}
\left({\dot a\over a}\right)^2 = {1\over 3 M_4^2} \rho - {k \over a^2}
+ {\Lambda_4 \over 3} + {1 \over 36 M_5^6}\rho^2 + {{\cal C} \over a^4},
\end{equation}
where $M_4 = (8 \pi G_N)^{-1/2}$ is the 4-dimensional Planck mass,
and $M_5$ is the 5-dimensional Planck mass.  In equation
(\ref{Friedmann}), the first three terms are identical
to the corresponding terms in the conventional Friedmann equation,
corresponding to the contribution from the total density, curvature,
and cosmological constant, respectively.  (The curvature and cosmological
constant have no effect on Big Bang nucleosynthesis, so we will not
discuss them further).  The final two terms give the modification
to the Friedmann equation in this brane scenario; the second of
these is sometimes called ``dark radiation", because it scales
as $a^{-4}$; however, ${\cal C}$ can be positive or negative.

Constraints can be placed on both $M_5$ and ${\cal C}$ from Big
Bang nucleosynthesis (BBN) \cite{Cline,Binetruy,Langlois,Barrow,Ichiki,Mizuno,
Maartens}.
These limits are based on the requirement that the change
in the expansion rate due to the quadratic and dark radiation terms
in equation (\ref{Friedmann}) be sufficiently small that an acceptable
$^4$He abundance be produced.  With the exception of Ref. \cite{Ichiki},
these limits are all rather rough.  In Ref. \cite{Ichiki}, the constraint
on the dark radiation term was calculated in detail using
limits from both BBN and the cosmic microwave background (CMB),
assuming negligible contribution from the quadratic term.
(For a further discussion of CMB limits on brane cosmologies, see
Ref. [12]).

In this paper, we extend the calculation of Ref. \cite{Ichiki}
to include both the contribution from the quadratic term and
the dark radiation simultaneously, providing a constraint in
the $M_5$, ${\cal C}$ plane.  This also provides the first detailed
calculation of the BBN limit on $M_5$ alone.   
Note that $M_5$ is
already constrained by the requirement that the theory reduce to
Newtonian gravity on scales $> 1$ mm; this requirement gives \cite{Maartens}
\begin{equation}
\label{M5limit}
M_5 > 10^5 ~{\rm TeV},
\end{equation}
which is considerably
stronger than limits that can be derived from BBN (which are
on the order of a few TeV).  However,
it has been noted \cite{Mizuno} that the type II Randall-Sundrum
model could be an effective theory, derived from a more fundamental
theory, in which case the limit in equation (\ref{M5limit}) need not apply.
It is therefore of interest to consider the BBN limit on $M_5$, both
alone and in combination with the limit on the dark radiation.

The primordial production of \4he is controlled by a competition between
the weak interaction rates (which govern the interconversion of neutrons
and protons) and the expansion rate of the Universe.  As long as the weak
interaction rates are faster than the expansion rate, the neutron-to-proton
ratio ($n/p$) tracks its equilibrium value.  Eventually, as the Universe expands and
cools, the expansion rate comes to dominate and $n/p$ essentially freezes out. 
The relatively large binding energy of \4he insures that nearly all the neutrons
which survive this freeze-out are converted into \4he as soon as deuterium
becomes stable against
photodisintegration
(see, e.g., Ref. \cite{Olive} for further details of this well-known story).  In the
standard cosmology [i.e., one where the Friedmann equation takes the form 
$({\dot a}/ a)^2 \sim \rho/(3 M_4^2)$], $n/p$ freezes out at 
a temperature $T \sim 1$ MeV.
Therefore, the primordial production of \4he is very sensitive to the expansion
rate of the Universe at temperatures $\sim$1 MeV, and in fact this sensitivity
has been exploited many times in bounding the number of light neutrino species.

Our analysis uses the sensitivity of primordial \4he to the expansion rate at 
$T \sim 1$ MeV 
to constrain the two additional ``brane-terms'' that appear in the type II 
Randall-Sundrum generalization of the Friedmann equation (equation 1). 
Specifically, we limit the additional energy density associated with the 
$M_5$ and dark radiation contributions by requiring that the primordial
production of \4he in such brane cosmologies agree with the observed abundance
of \4he.  Note that the primordial
production of \4he is weakly (log) dependent on the baryon density and so
in principle constraints on such brane cosmologies would also involve
constraints on the baryon density.
However, unlike \4he, the
abundance of deuterium is very sensitive to the baryon density and fairly
insensitive to the expansion rate (which we verify in our detailed
calculations).
Therefore, we can use a comparison of the predicted deuterium abundance and its
observed abundance to independently constrain the baryon density and then 
examine the subsequent constraints on $M_5$ and the dark radiation coming
from \4he.

In order to constrain the additional brane-terms, we start with a 
standard BBN model and
some conservative estimates of the primordial abundances of the light elements.
We assume 3 light neutrino species and the standard updated nuclear reaction
network.
Because the dark radiation scales as $a^{-4}$, it is convenient
to parametrize it in terms of the effective
number of additional neutrino species, $\Delta N_\nu$, given by
$\Delta N_\nu \equiv ({\cal C}/ a^4)(3 M_4^2/\rho_\nu)$,
where $\rho_\nu$ is the energy density contributed by
a single, two-component massless neutrino, and $\Delta N_\nu$ can
be either positive or negative.
(Note that a slightly
different parametrization was used in Ref. \cite{Ichiki}).
Then the brane Friedman equation can be completely specified by
$M_5$ and $\Delta N_\nu$.

We take the primordial deuterium abundance as inferred from
QSO absorption line systems in the range 
\begin{equation}
3\times 10^{-5}< {\rm D/H} < 4\times
10^{-5},
\end{equation}
and we take Y$_p$ (the primordial mass fraction of \4he)
as inferred from low-metallicity HII regions to be
\begin{equation}
0.23 < {\rm Y}_p < 0.25,
\end{equation}
(see, e.g., Ref. \cite{Olive}).
For a fixed pair of $M_5$ and $\Delta N_\nu$, we then scan
over the baryon-photon ratio $\eta$ to see if a value of $\eta$
exists which produces deuterium and $^4$He abundances within
the acceptable limits.  In practice, the deuterium abundance
is nearly insensitive to $M_5$ and $\Delta N_\nu$, so the deuterium
constraint limits $\eta$ to the range $\eta = 4-6 \times 10^{-10}$
for most of our parameter range,
and the upper and lower limits on $^4$He then give the contraints
in the $M_5$ -- dark radiation plane (although note that our procedure
is ``exact" in the sense described above; we do not assume a fixed
bound on $\eta$).
Our results are shown in Fig. 1, where we
indicate the region in the $M_5$ -- dark radiation plane consistent with our
adopted range of primordial abundances. (All the brane-cosmologies in our
allowed region produce acceptable
primordial $^7$Li abundances.)  

\begin{figure}
\begin{center}
\epsfxsize=3.5truein
\epsfysize=3.5truein
\epsfbox{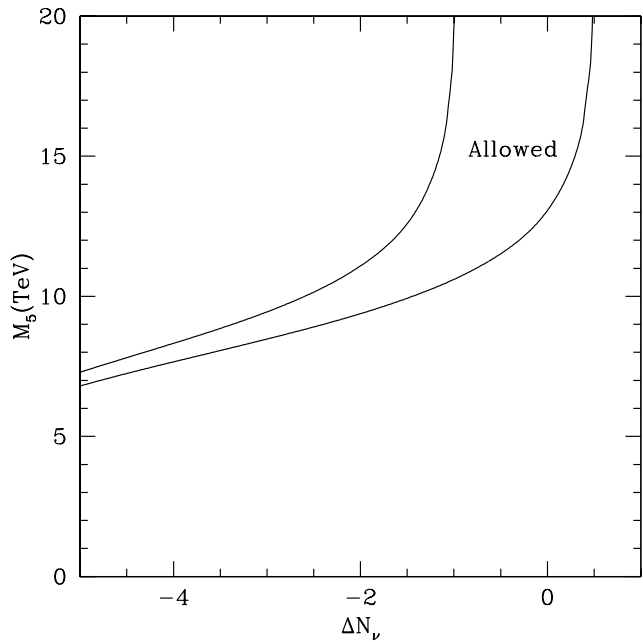}
\end{center}
\baselineskip 7pt
\caption{The area between the two curves gives the region in the $M_5$--dark radiation
plane allowed
by Big Bang nucleosynthesis, where $M_5$ is the 5-dimensional Planck mass,
and the dark radiation is parametrized in terms of $\Delta N_\nu$, the
effective number of additional two-component neutrinos.  In terms
of the quantities appearing in the brane-cosmology Friedman equation,
${\cal C}/a^4 = (\Delta N_\nu \rho_\nu)/(3 M_4^2)$, where $\rho_\nu$ is the energy
density of a single two-component neutrino.}
\label{fig2}
\end{figure}

We can understand the shape of the excluded region shown in Fig. 1 as follows.
The left and right contours of the allowed region correspond to
Y$_p = 0.23$ (lowest allowed expansion rate) and Y$_p = 0.25$ (largest
allowed expansion rate), respectively.  
The sensitivity of primordial \4he
to the expansion
of the Universe at $T \sim 1$ MeV constrains the additional energy
density at that temperature
to be within roughly half a neutrino species of the standard model.
When $M_5$ is sufficiently large, we get
the standard constraint on additional relativistic energy density,
which appears in Fig. 1 as a vertical band.
As $M_5$ decreases,
the $\rho^2$ term in the Friedmann equation starts to dominate at 
$T \sim 1$ MeV and our
allowed region moves sharply to the left, so that the negative dark
radiation energy density cancels the $\rho^2$ contribution to
the expansion.

Our investigation yields two new results:  a precision determination
of the BBN limit on $M_5$, and a combined limit on $M_5$ and ${\cal C}$.
For ${\cal C} = 0$,
we derive the precise BBN limit on $M_5$ alone:
\begin{equation}
M_5 > 13 ~{\rm TeV}.
\end{equation}
Our limits on the dark radiation for large $M_5$ are
\begin{equation}
-1.0 < \Delta N_\nu < 0.5,
\end{equation}
in rough agreement with Ref. \cite{Ichiki}. 
When both the quadratic term and the dark radiation contribute
significantly to the Friedman equation, our limits are given
by the contours in Fig. 1.  Although we can find acceptable
solutions for very small $M_5$, the allowed region shrinks
dramatically in this limit; a small $M_5$ requires both ${\cal C} < 0$
and precise
fine-tuning of ${\cal C}$.

\acknowledgments
J.B. and A.C.G. were supported at Ohio State University
under the NSF Research Experience for
Undergraduates (REU) program (PHY-9912037).
R.J.S. and T.P.W. were supported by the Department
of Energy (DE-FG02-91ER40690).

\newcommand\AJ[3]{~Astron. J.{\bf ~#1}, #2~(#3)}
\newcommand\APJ[3]{~Astrophys. J.{\bf ~#1}, #2~ (#3)}
\newcommand\apjl[3]{~Astrophys. J. Lett. {\bf ~#1}, L#2~(#3)}
\newcommand\ass[3]{~Astrophys. Space Sci.{\bf ~#1}, #2~(#3)}
\newcommand\cqg[3]{~Class. Quant. Grav.{\bf ~#1}, #2~(#3)}
\newcommand\mnras[3]{~Mon. Not. R. Astron. Soc.{\bf ~#1}, #2~(#3)}
\newcommand\mpla[3]{~Mod. Phys. Lett. A{\bf ~#1}, #2~(#3)}
\newcommand\npb[3]{~Nucl. Phys. B{\bf ~#1}, #2~(#3)}
\newcommand\plb[3]{~Phys. Lett. B{\bf ~#1}, #2~(#3)}
\newcommand\pr[3]{~Phys. Rev.{\bf ~#1}, #2~(#3)}
\newcommand\PRL[3]{~Phys. Rev. Lett.{\bf ~#1}, #2~(#3)}
\newcommand\PRD[3]{~Phys. Rev. D{\bf ~#1}, #2~(#3)}
\newcommand\prog[3]{~Prog. Theor. Phys.{\bf ~#1}, #2~(#3)}
\newcommand\RMP[3]{~Rev. Mod. Phys.{\bf ~#1}, #2~(#3)}

\end{document}